\documentclass[12pt,draft]{iopart}
\usepackage{amsfonts}
\usepackage[english]{babel}
\usepackage{euscript}
\usepackage{bbm}

       \newcommand{\Ec}{\EuScript{E}}
       \newcommand{\Hc}{\EuScript{H}}
       \newcommand{\Vc}{\EuScript{V}}
       \newcommand{\N}{\mathbb{N}}
       \newcommand{\R}{\mathbb{R}}
       
\begin{document}

\title[Symmetry group and a covering for modified Khokhlov-Zabolotskaya equation]%
{Cartan's structure of symmetry pseudo-group and a covering for the modified 
Khokhlov-Zabolotskaya equation}

\author{Oleg I. Morozov}

\address{Department of Mathematics, Moscow State Technical University 
\\
of Civil Aviation, Kronshtadtskiy Blvd 20, Moscow 125993, Russia
\\
oim{\symbol{64}}foxcub.org}

\begin{abstract}
We apply Cartan's method of equivalence to find a covering for the mo\-di\-fi\-ed Khokhlov--Zabolotskaya equation.
\end{abstract}


\ams{58H05, 58J70, 35A30}

\section{Introduction}
In this paper we derive  a covering for the modified Khokhlov-Zabolotskaya equation \cite{Kuzmina} (or the dispersionless modified Kadomtsev-Petviashvili equation, \cite{ChangTu})
\begin{equation}
u_{yy} = u_{tx} +\left(\case{1}{2}\,u_x^2-u_y\right)\,u_{xx}
\label{mKhZ}
\end{equation}
from Maurer-Cartan forms (MC forms) of its symmetry pseudo-group.

Coverings \cite{KV84,KV89,KV99}  (or prolongation structures \cite{WE}, or zero-cur\-va\-tu\-re re\-pre\-sen\-ta\-tions \cite{ZakharovShabat}, or integrable extensions \cite{BryantGriffiths}) are of great im\-por\-tan\-ce in geometry of dif\-fe\-ren\-ti\-al equations. They are a starting point for inverse scattering transformations, B\"ack\-lund transformations, recursion operators, nonlocal symmetries and nonlocal conservation laws. Different techniques are developed for constructing coverings of partial differential equations (PDEs) in two independent va\-ri\-ables, \cite{WE,DoddFordy,Estabrook,Hoenselaers,Marvan,Marvan2002,Sakovich,Igonin}, while in the case of more than two independent va\-ri\-ables the problem is more difficult, see, e.g.,  \cite{Morris1976,Morris1979,Zakharov82,Tondo,Marvan,Harrison1995,Harrison2000}.  In the pioneering work \cite{Kuzmina}, Cartan's method of equivalence was applied to the covering problem for equations in three independent variables. One of the results of \cite{Kuzmina} is a deduction of the system 
\begin{eqnarray}
v_t = (v^2-u)\,v_x - u_y - v\,u_x,
\label{KC1}
\\
v_y = v\,v_x - u_x,
\label{KC2}
\end{eqnarray}
whose integrability conditions coincide with the Khokhlov-Zabolotskaya equation \cite{KhZ}
\begin{equation}
u_{yy} = u_{tx} + u\,u_{xx} + u_x^2.
\label{KhZ}
\end{equation}
In terms of \cite{KV84,KV89,KV99}, system (\ref{KC1}), (\ref{KC2}) defines an infinite-dimensional covering for equa\-ti\-on (\ref{KhZ}). Also, a B\"acklund transformation from (\ref{KhZ}) to (\ref{mKhZ}) is found in \cite{Kuzmina}: from (\ref{KC2}) it fol\-lows that there exists 
a function $w$ such that $w_x = v$ and $w_y = \case{1}{2}\,v^2 - u$; then from (\ref{KC1}) it follows that $w$ satisfies 
(\ref{mKhZ}).

The present paper is an attempt to clarify the method of \cite{Kuzmina}. We apply Cartan's method of equivalence, \cite{Cartan1}--\cite{Cartan5}, \cite{Gardner,Kamran,Olver95,FelsOlver}, to compute MC forms for the pseudo-group of contact symmetries of (\ref{mKhZ}) and then find their linear combination that gives covering equations for (\ref{mKhZ}).

\section{Cartan's structure theory of contact symmetry pseudo-groups of DEs}

In this section, we outline the algorithm of computing MC forms for pseudo-groups of contact symmetries for DEs of the second order with one dependent variable, see details in \cite{Morozov2002,Morozov2006}. All considerations are of local nature, and all mappings are real analytic. Let $\pi :\R^n \times \R \rightarrow \R^n$ be a vector bundle with the local base co\-or\-di\-na\-tes $(x^1,...,x^n)$ and the local fibre coordinate $u$; then by $J^2(\pi)$ denote the bundle of the second-order jets of sections of $\pi$, with the local coordinates $(x^i,u,u_i,u_{ij})$, $i,j\in\{1,...,n\}$, $i \le j$. For every local section $(x^i,f(x))$ of $\pi$, denote by $j_2(f)$ the corresponding 2-jet $(x^i,f(x),\partial f(x)/\partial x^i,\partial^2 f(x)/\partial x^i\partial x^j)$. A differential 1-form $\vartheta$ on $J^2(\pi)$ is called a {\it contact form} if it is annihilated by all 2-jets of local sections: $j_2(f)^{*}\vartheta = 0$. In the local coordinates every contact 1-form is a linear combination of the forms  $\vartheta_0 = du - u_{i}\,dx^i$,
$\vartheta_i = du_i - u_{ij}\,dx^j$, $i, j \in \{1,...,n\}$, $u_{ji} = u_{ij}$ (here and later we use the Einstein summation convention, so $u_i\,dx^i = \sum_{i=1}^{n}\,u_i\,dx^i$, etc.) A local dif\-feo\-mor\-phism $\Delta : J^2(\pi) \rightarrow J^2(\pi)$, 
$\Delta : (x^i,u,u_i,u_{ij}) \mapsto (\bar{x}^i,\bar{u},\bar{u}_i,\bar{u}_{ij})$,
is called a {\it contact trans\-for\-ma\-tion} if for every contact 1-form $\bar{\vartheta}$ the form $\Delta^{*}\bar{\vartheta}$ is also contact. We denote by $\rm{Cont}(J^2(\pi))$  the pseudo-group  of contact transformations on $J^2(\pi)$.

Let $\Hc$ be a open subset of $\R^{(2 n+1)(n+3)(n+1)/3}$ with local coordinates
$(a$, $b^i_k$, $c^i$, $f^{ik}$, $g_i$, $s_{ij}$, $w^k_{ij}$, $z_{ijk})$, $i,j,k \in \{1,...,n\}$, 
$i \le j$, such that $a\not =0$, $\det (b^i_k) \not = 0$, $f^{ik}=f^{ki}$ and $z_{ijk}=z_{ikj}=z_{jik}$. Let $(B^i_k)$ be the inverse matrix for the matrix $(b^k_l)$, so
$B^i_k\,b^k_l = \delta^i_l$. We consider the {\it lifted coframe}
\[
\Theta_0 = a\, \vartheta_0,
\quad
\Theta_i = g_i\,\Theta_0 + a\,B_i^k\,\vartheta_k,
\quad
\Xi^i =c^i\,\Theta_0+f^{ik}\,\Theta_k+b_k^i\,dx^k,
\]
\begin{equation}
\Sigma_{ij} = s_{ij}\,\Theta_0+w_{ij}^{k}\,\Theta_k+z_{ijk}\,\Xi^k + a\,B^i_k\, B^j_l\,du_{kl},
\label{LCF}
\end{equation}
\noindent
defined on $J^2(\pi)\times\Hc$. As it is shown in \cite{Morozov2006}, the forms (\ref{LCF}) are MC forms for $\rm{Cont}(J^2(\pi))$, that is, a local diffeomorphism
$\widehat{\Delta} : J^2(\pi) \times \Hc \rightarrow J^2(\pi) \times \Hc$
satisfies the conditions
$\widehat{\Delta}^{*}\, \bar{\Theta}_0 = \Theta_0$,
$\widehat{\Delta}^{*}\, \bar{\Theta}_i = \Theta_i$,
$\widehat{\Delta}^{*}\, \bar{\Xi}^i = \Xi^i$,
and $\widehat{\Delta}^{*}\, \bar{\Sigma}_{ij} = \Sigma_{ij}$
if and only if it is projectable on $J^2(\pi)$, and its projection
$\Delta : J^2(\pi) \rightarrow J^2(\pi)$ is a contact transformation.

The structure equations for $\rm{Cont}(J^2(\pi))$ have the form
\[
d \Theta_0 = \Phi^0_0 \wedge \Theta_0 + \Xi^i \wedge \Theta_i,
\]
\[
d \Theta_i = \Phi^0_i \wedge \Theta_0 + \Phi^k_i \wedge \Theta_k
+ \Xi^k \wedge \Sigma_{ik},
\]
\[
d \Xi^i = \Phi^0_0 \wedge \Xi^i -\Phi^i_k \wedge \Xi^k
+\Psi^{i0} \wedge \Theta_0
+\Psi^{ik} \wedge \Theta_k,
\]
\[
d \Sigma_{ij} = \Phi^k_i \wedge \Sigma_{kj} - \Phi^0_0 \wedge \Sigma_{ij}
+ \Upsilon^0_{ij} \wedge \Theta_0
+ \Upsilon^k_{ij} \wedge \Theta_k + \Lambda_{ijk} \wedge \Xi^k,
\]
\noindent
where the additional forms $\Phi^0_0$, $\Phi^0_i$, $\Phi^k_i$, $\Psi^{i0}$, $\Psi^{ij}$,
$\Upsilon^0_{ij}$, $\Upsilon^k_{ij}$, and $\Lambda_{ijk}$ depend on differentials of the coordinates of $\Hc$.

Suppose $\Ec$ is a second-order differential equation in one dependent and
$n$ independent variables. We consider $\Ec$ as a submanifold in $J^2(\pi)$.
Let $\rm{Cont}(\Ec)$ be the group of contact symmetries for $\Ec$. It consists of
all the contact transformations on $J^2(\pi)$ mapping $\Ec$ to itself.
Let $\iota_0 : \Ec \rightarrow J^2(\pi)$ be an embedding, and 
$\iota = \iota_0 \times \rm{id} : \Ec\times \Hc \rightarrow J^2(\pi)\times \Hc$.
The invariant 1-forms of $\rm{Cont}(\Ec)$ are re\-stric\-ti\-ons of the forms (\ref{LCF})
to $\Ec\times \Hc$: $\theta_0 = \iota^{*} \Theta_0$, $\theta_i= \iota^{*}\Theta_i$,
$\xi^i = \iota^{*}\Xi^i$, and $\sigma_{ij}=\iota^{*}\Sigma_{ij}$.
The forms $\theta_0$, $\theta_i$, $\xi^i$, and $\sigma_{ij}$ have some linear dependencies, 
i.e., there exists a non-trivial set of functions $E^0$, $E^i$, $F_i$, and $G^{ij}$ on
$\Ec\times \Hc$ such that  
$E^0\,\theta_0 + E^i\,\theta_i + F_i\,\xi^i+ G^{ij}\,\sigma_{ij} \equiv 0$.
These functions are lifted invariants of $\rm{Cont}(\Ec)$. Setting them equal to
some constants allows us to specify some coordinates $a$, $b^k_i$, $c_i$,
$g_i$, $f^{ij}$, $s_{ij}$, $w^k_{ij}$, and $z_{ijk}$
as functions of the coordinates on $\Ec$ and the other coordinates on $\Hc$.

After these normalizations, a part of the forms $\phi^0_0=\iota^{*}\Phi^0_0$,
$\phi^k_i=\iota^{*}\Phi^k_i$, $\phi^0_i=\iota^{*}\Phi^0_i$,
$\psi^{ij}=\iota^{*}\Psi^{ij}$, $\psi^{i0}=\iota^{*}\Psi^{i0}$,
$\upsilon^{0}_{ij}=\iota^{*}\Upsilon^{0}_{ij}$,
$\upsilon^{k}_{ij}=\iota^{*}\Upsilon^{k}_{ij}$, and
$\lambda_{ijk}=\iota^{*}\Lambda_{ijk}$, or some their li\-ne\-ar combinations,
become semi-basic, i.e., they do not include the differentials of the
coordinates on $\Hc$.
Setting coefficients of the semi-basic forms  equal to some constants, we get specifications of some
more coordinates on $\Hc$.

More lifted invariants can appear as essential torsion coefficients in the
reduced structure equations
\[
d \theta_0 = \phi^0_0 \wedge \theta_0 + \xi^i \wedge \theta_i,
\]
\[
d \theta_i =
\phi^0_i \wedge \theta_0 + \phi^k_i \wedge \theta_k + \xi^k \wedge \sigma_{ik},
\]
\[
d \xi^i = \phi^0_0 \wedge \xi^i -\phi^i_k \wedge \xi^k +\psi^{i0} \wedge \theta_0
+\psi^{ik} \wedge \theta_k,
\]
\[
d \sigma_{ij} = \phi^k_i \wedge \sigma_{kj} - \phi^0_0 \wedge \sigma_{ij}
+ \upsilon^0_{ij} \wedge \theta_0 + \upsilon^k_{ij} \wedge \theta_k
+ \lambda_{ijk} \wedge \xi^k.
\]
\noindent
After normalizing these invariants and repeating the process, two outputs are
possible. In the first case, the reduced lifted coframe appears to be
involutive. Then this coframe is the desired set of MC forms for
$\rm{Cont}(\Ec)$. In the second case, when the reduced lifted coframe does not
satisfy Cartan's test, we should use the procedure of prolongation,
\cite[ch~12]{Olver95}.

\section{Coverings of DEs}

Let $\pi_{\infty} : J^{\infty}(\pi) \rightarrow \R^n$ be the infinite jet bundle of local sections of the bundle $\pi$. The coordinates on $J^{\infty}(\pi)$ are $(x^i, u, u_I)$, where
$I = (i_1,...,i_k)$ are symmetric multi-indices,  $i_1,...,i_k \in \{1,...,n\}$, and for any local section $f$ of $\pi$ there exists a section $j_{\infty}(f) : \R^n \rightarrow J^{\infty}(\pi)$
such that $u_I(j_{\infty}(f)) = \partial^{\# I}(f)/\partial x^{i_1} ... \partial x^{i_k}$,
$\# I =\#(i_1,...,i_k) = k$.  Contact forms on $J^{\infty}(\pi)$ are defined by the requirement to satisfy $j_{\infty}(f)^{*}\,\vartheta = 0$ for any $f$. They are linear combinations
of the forms $\vartheta_I = d u_I - u_{Ii}\,d x^i$, $\# I \ge 0$. The {\it total derivatives} on
$J^{\infty}(\pi)$ are defined in the local coordintes as
\[
D_i = \frac{\partial}{\partial x^i}
+\sum \limits_{\# I \ge 0} u_{Ii} \, \frac{\partial}{\partial u_I}.
\]
\noindent
We have $[D_i, D_j] = 0$ for $i, j \in \{1,...,n\}$ and
$\vartheta_I = D_I (\vartheta_0)$, where $D_I = D_{i_1}\circ ... \circ D_{i_k}$ for $I=(i_1,...,i_k)$.

A differential equation $F(x^i, u, u_I)=0$,  $\#I \le q$,  defines a submanifold
\[
\Ec^{\infty} = \{ D_K(F) =0 \,\,\vert\,\, \#K\ge 0\} \subset J^{\infty}(\pi).
\]

\noindent
We denote restrictions of $D_i$ and $\vartheta_I$ on $\Ec^{\infty}$ as
$\bar{D}_i$ and $\bar{\vartheta}_I$, respectively.

In local coordinates, a {\it covering} over $\Ec^{\infty}$ is a bundle
$\widetilde{\Ec}^{\infty} = \Ec^{\infty} \times \Vc \rightarrow \Ec^{\infty}$ with fibre coordinates $v^\kappa$, $\kappa \in \{1,..., N\}$ or $\kappa \in \N$, equipped with ex\-ten\-ded total derivatives
\[
\widetilde{D}_i = \bar{D}_i
+\sum \limits_{\kappa}
T^{\kappa}_i (x^j, u, u_I, v^{\tau})\,\frac{\partial}{\partial v^\kappa},
\qquad i \in \{1,...,n\},
\]

\noindent
such that $[\widetilde{D}_i, \widetilde{D}_j ]=0$ whenever $(x^i,u,u_I) \in \Ec^{\infty}$.

In terms of differential forms, the covering is defined by the forms
\[
\widetilde{\vartheta}^{\kappa} = d v^{\kappa}
- T^{\kappa}_i (x^j, u, u_I, v^{\tau})\,dx^i
\]
\noindent
such that
$d \widetilde{\vartheta}^{\kappa} \equiv 0 \,\,\,({\mathrm{mod}}\,\,\,
\widetilde{\vartheta}^{\tau}, \bar{\vartheta}_I )$ whenever $(x^i,u,u_I) \in \Ec^{\infty}$.
We call $\widetilde{\vartheta}^{\kappa}$ {\it Wahlquist--Estabrook forms} (WE forms) of the covering.
\vskip 5 pt
\noindent
EXAMPLE. 
System (\ref{KC1}), (\ref{KC2}) provides an infinite-dimensional covering for (\ref{KhZ}) with fibre coordinates $v_0=v$, $v_k = \partial^k\,v/\partial x^k$, $k \in \mathbb{N}$, the extended total derivatives 
\begin{eqnarray*}
\widetilde{D}_t = \bar{D}_t 
+ \sum \limits_{j =0}^{\infty} \widetilde{D}^j_x((v_0^2-u)\,v_1-u_y-v_0\,u_x)\,
\frac{\partial}{\partial v_j},
\\
\widetilde{D}_x = \bar{D}_x + \sum \limits_{j =0}^{\infty} v_{j+1}\,\frac{\partial}{\partial v_j},
\\
\widetilde{D}_y = \bar{D}_y 
+ \sum \limits_{j =0}^{\infty} \widetilde{D}^j_x(v_0\,v_1-u_x)\,\frac{\partial}{\partial v_j},
\end{eqnarray*}
and the WE forms
\begin{eqnarray*}
\widetilde{\vartheta}_0 = d v_0 - ((v_0^2-u)\,v_1-u_y-v_0\,u_x)\,dt - v_1\,dx
- (v_0\,v_1-u_x)\,dy,
\\
\widetilde{\vartheta}_{k} = \widetilde{D}^k_x(\widetilde{\vartheta}_0),
\qquad k \in \N.
\end{eqnarray*}

\section{Symmetry pseudo-group and a covering for the modified Khokhlov - Zabolotskaya equation}

By the method described in section 2 we compute MC forms and structure equations for the pseudo-group of contact symmetries of equation (\ref{mKhZ}). The structure equations read
\begin{eqnarray*} 
\fl
d\theta_{0}=\eta_1 \wedge \theta_{0}+
\xi^1 \wedge \theta_1+\xi^2 \wedge \theta_2+\xi^1 \wedge \theta_2+\xi_
3 \wedge \theta_3,
\\
\fl
d\theta_1=
\left(\case{1}{2}\,\theta_2+\xi^2\right) \wedge \theta_{0}
+\left(\case{3}{2}\,\eta_1 +\xi^3 -\case{3}{2}\,\sigma_{22} \right)\wedge \theta_1
+\left(\eta_1 +\theta_3 -\sigma_{22} +\xi^3\right) \wedge \theta_2
+2\,\theta_3 \wedge \xi^2 
\\ 
\fl \qquad
+\xi^1 \wedge \sigma_{11}
+(\xi^1+\xi^2) \wedge \sigma_{12}
+\xi^3 \wedge \sigma_{13},
\\
\fl
d\theta_2=
\case{1}{2}\,(\eta_1 - \sigma_{22}) \wedge \theta_2
+\xi^1 \wedge \sigma_{12}
+(\xi^1 +\xi^2) \wedge \sigma_{22} 
+\xi^3 \wedge \sigma_{23},
\\
\fl
d\theta_3=
\case{1}{2}\,\sigma_{22} \wedge \theta_{0}
-\xi^2 \wedge \theta_2
+\left(\eta_1 +\case{1}{2}\,\xi^3 -\sigma_{22}\right) \wedge \theta_3
+\xi^1 \wedge \sigma_{13}
+\xi^3 \wedge (\sigma_{12}+\sigma_{22})
\\ 
\fl \qquad
+(\xi^1  +\xi^2) \wedge \sigma_{23},
\\
\fl
d\xi^1 =-\case{1}{2}\,\left(\eta_1 +2\,\xi^3 -3\,\sigma_{22}\right) \wedge \xi^1,
\\
\fl
d\xi^2=
\left(\theta_3 -\case{1}{2}\,\theta_{0}\right) \wedge \xi^1
+\case{1}{2}\,(\eta_1 +\sigma_{22}) \wedge \xi^2 
+(\theta_2+\xi^2) \wedge \xi^3,
\\
\fl
d\xi^3=
2\,(\theta_2 +\xi^2) \wedge \xi^1
+\sigma_{22} \wedge \xi^3,
\\
\fl
d\sigma_{11}=
2\,\eta_1 \wedge (\sigma_{11}+\sigma_{12})
+\eta_2 \wedge \xi^1
+\eta_3 \wedge (\xi^1+\xi^2) 
+\eta_4 \wedge \xi^3
+\sigma_{22} \wedge \theta_{0}
\\ 
\fl \qquad
+3\,(2\,\theta_2-\sigma_{23}) \wedge \theta_1
+ (3\,\sigma_{13}-2\,\sigma_{23}) \wedge \theta_2
+(\theta_3+3\,\sigma_{11}+2\,\sigma_{12}) \wedge \sigma_{22},
\\
\fl
d\sigma_{12}=
\eta_1 \wedge (\sigma_{12}+\sigma_{22})
+\eta_3 \wedge \xi^1
+\eta_5 \wedge \xi^3
+\case{1}{2}\,\theta_{0} \wedge \sigma_{22}
+\case{13}{2}\,\theta_1 \wedge \xi^1
\\
\fl\qquad
+\case{1}{2}\,\theta_2 \wedge (11\,\xi^1+3\,\xi^2+2\,\sigma_{23})
-\theta_3 \wedge \sigma_{22} 
-2\,(2\,\sigma_{13}+\sigma_{23}) \wedge \xi^1
+2\,\sigma_{12} \wedge \sigma_{22},
\\
\fl
d\sigma_{13}=
\case{1}{2}\,\eta_1 \wedge (3\,\sigma_{13}+\sigma_{23})
+\eta_3 \wedge \xi^3
+\eta_4 \wedge \xi^1
+\eta_5 \wedge (\xi^1+\xi^2)
+\case{1}{2}\,\theta_{0} \wedge (3\,\theta_2 +2\, \xi^2 -\sigma_{23})
\\
\fl\qquad
+\case{1}{2}\,\theta_1 \wedge (13\,\xi^3-\sigma_{22})
+\case{1}{2}\,\theta_2 \wedge (6\,\theta_3 +13\,\xi^3-4\,\sigma_{12} -2\,\sigma_{22})
-\theta_3 \wedge (4\,\xi^2-\sigma_{23}) 
\\
\fl\qquad
+(2\,\sigma_{11}+3\,\sigma_{12} +\sigma_{22}) \wedge \xi^1
+(4\,\sigma_{12}+3\,\sigma_{22}) \wedge \xi^2
+\case{1}{2}\,xi^3 \,\wedge (11\,\sigma_{13}+6\,\sigma_{23})
\\
\fl\qquad
-\case{1}{2}\,\sigma_{22} \wedge (5\,\sigma_{13} +2\,\sigma_{23}),
\\
\fl
d\sigma_{22}=
2\,(2\,\theta_2 +2\,\xi^2-\sigma_{23}) \wedge \xi^1
+\case{1}{2}\sigma_{22} \wedge \xi^3,
\\
\fl
d\sigma_{23}=
\case{1}{2}\,\eta_1 \wedge \sigma_{23}
+\eta_5 \wedge \xi^1
+\case{3}{2}\,\theta_2 \wedge (2\,\xi^3-\sigma_{22})
+\case{1}{2}\,\xi^1 \wedge (3\,\sigma_{22} -2\,\xi^3-2\,\sigma_{12})
\\
\fl\qquad
+\case{3}{2}\,\xi^2 \wedge (2\,\xi^3-\sigma_{22})
+\case{1}{2}\,(5\,\xi^3 -3\,\sigma_{22}) \wedge \sigma_{23},
\\
\fl
d\eta_1=\xi^1 \wedge (\theta_2 +\xi^2) \wedge \xi^1 +\case{1}{2}\,\xi^3 \wedge \sigma_{22},
\\
\fl
d\eta_2=\pi_1 \wedge \xi^1+\pi_2 \wedge (\xi^1+\xi^2) +\pi_3 \wedge \xi^3
+\case{1}{2}\,
\eta_1 \wedge (5\,\eta_2+6\,\eta_3 -13\,\theta_1+16\,\theta_2-16\,\sigma_{13} 
\\
\fl \qquad
-8\,\sigma_{23})
+\case{9}{2}\,\eta_2 \wedge \sigma_{22}
\case{1}{2}\,\eta_3 \wedge (2\,\theta_3-\theta_{0}+6\,\sigma_{22})
+5\,\eta_4 \wedge \theta_2
-\eta_5 \wedge (3\,\theta_1+2\,\theta_2)
\\
\fl \qquad
-\theta_{0} \wedge (16\,\theta_2-5\,\sigma_{23}) 
+\theta_1 \wedge (9\,\sigma_{12}+14\,\sigma_{22})
+(32\,\theta_3+26\,\sigma_{11}+5\,\sigma_{12}-12\,\sigma_{22}) \wedge \theta_2
\\
\fl \qquad
+10\,\sigma_{23} \wedge \theta_3
-3\,(\sigma_{13}-2\,\sigma_{23}) \wedge \sigma_{12}
+\sigma_{23} \wedge (9\,\sigma_{11}-4\,\sigma_{22})
+10\,\sigma_{22} \wedge \sigma_{13},
\\
\fl
d\eta_3=\pi_2 \wedge \xi^1+\pi_4 \wedge \xi^3
+\case{1}{2}\,\eta_1 \wedge (6\,\theta_2 +8\,\xi^1+8\,\xi^2+3\,\eta_3)
+\case{7}{2}\,\eta_3 \wedge \sigma_{22}
-2\,\eta_4 \wedge \xi^1
\\
\fl \qquad
+3\,\eta_5 \wedge (\theta_2+2\,\xi^1 +2\,\xi^2)
+\case{1}{2}\,\theta_0 \wedge (24\,\theta_2+41\,\xi^1 
+33\,\xi^2-3\,\sigma_{23})
+\case{21}{2}\,\theta_1 \wedge \sigma_{22}
\\
\fl \qquad
+6\,\theta_2 \wedge (4\,\theta_3-3\,\sigma_{12})
-\theta_3 \wedge (41\,\xi^1 -33\,\xi^2-6\,\sigma_{23}) 
+14\,\xi^1 \wedge \sigma_{11}
+20\,\sigma_{22} \wedge (\xi^1+\xi^2)
\\
\fl \qquad
+10\,\sigma_{12} \wedge (2\,\xi^1+3\,\xi^2)
-3\,\sigma_{12} \wedge \sigma_{23}
-3\,\sigma_{13} \wedge \sigma_{22},
\\ 
\fl
d\eta_4 =
\pi_2 \wedge \xi^3
+\pi_3 \wedge \xi^1
+\pi_4 \wedge (\xi^1+\xi^2) 
+2\,\eta_1 \wedge (\eta_4 +\eta_5 +2\,\xi^3)
+\eta_2 \wedge \xi^1
\\
\fl \qquad
+\eta_3 \wedge (4\,\theta_2+\xi^1+\xi^2)
+\case{1}{2}\,\eta_4 \wedge (8\,\sigma_{22}-21\,\xi^3) 
+2\,\eta_5 \wedge (\sigma_{22}-\xi_3)
+\case{1}{2}\,\theta_{0} \wedge (43\,\xi^3-\sigma_{22})
\\
\fl \qquad
+\case{1}{2}\,\theta_1 \wedge (69\,\theta_2-3\,\xi^1+9\,\xi^2-9\,\sigma_{23})
+\case{1}{2}\,\theta_2 \wedge (4\,\xi^1 + 12\,\xi^2+21\,\sigma_{13}+\sigma_{23})
\\
\fl \qquad
-\theta_3 \wedge (43\,\xi^3-\sigma_{22})
+\case{1}{2}\,\xi^3 \wedge (67\,\sigma_{11}+18\,\sigma_{12}-28\,\sigma_{22})
+\case{5}{2}\,\sigma_{11} \wedge \sigma_{22}
-6\,\sigma_{13} \wedge \sigma_{23},
\\
\fl
d\eta_5=
\pi_4 \wedge \xi^1
+\case{1}{2}\,\eta_1 \wedge (2\,\eta_5+2\,\xi^3+\sigma_{22})
-3\,\eta_3 \wedge \xi^1
+\case{3}{2}\,\eta_5 \wedge (2\,\sigma_{22}-2\,\xi^3)
\\
\fl \qquad
-\case{1}{4}\,\theta_{0} \wedge (6\,\xi^3 -\sigma_{22})
-26\,\theta_1 \wedge \xi^1
+\case{1}{2}\,\theta_2 \wedge (5\,\sigma_{23}-64\,\xi^1-3\,\xi^2)
+\case{1}{2}\,\theta_3 \wedge (6\,\xi^3 -\sigma_{22})
\\
\fl \qquad
+\case{1}{2}\,\xi^1 \wedge (12\,\xi^2-23\,\sigma_{13} -19\,\sigma_{23})
+\case{3}{2}\,\xi^2 \wedge \sigma_{23}
+\case{13}{2}\,\xi^3 \wedge (\sigma_{12}+2\,\sigma_{22})
+\case{5}{2}\,\sigma_{12} \wedge \sigma_{22}.
\end{eqnarray*}
The forms $\eta_1$, ... , $\eta_5$ appear in the step of absorption of torsion in the reduced structure equations. We have
\begin{eqnarray*}
\xi^1=q\,dt,
\\
\xi_2=u_{xx}^2\,q^{-1}\,
\left(dx+ u_x\,dy+\left(\case{1}{2}\,u_x^2+u_y\right)\,dt\right),
\\
\xi_3=u_{xx} \,(2\,u_x\,dt+dy),
\\
\eta_1=3\,(u_{xx})^{-1}\,du_{xx}-2\,q^{-1}\,dq-\case{1}{2}\,u_{xx}\,dy-u_x\,u_{xx}\,dt,
\end{eqnarray*}
with $q=b^1_1 \not =0$. We need not explicit expressions for the other MC forms in the sequel. We take a linear combination
\[\fl\qquad
\eta_1+\xi^2+\case{1}{2}\,\xi^3=
3\,(u_{xx})^{-1}\,du_{xx}-2\,q^{-1}\,dq
+\case{1}{2}\,u_{xx}^2\,
\left(\left(\case{1}{2}\,u_x^2+u_y\right)\,dt 
+dx
+u_x\,dy
\right)
\]
and substitute $u_{xx}= v^2\,v_1^2$, $q=\case{1}{4}\,v^5\,v_1^3$. Then we have
\[
\eta_1+\xi^2+\case{1}{2}\,\xi^3 = 
-4\,v^{-1}\,  
\left(dv-\left(\case{1}{2}\,u_x^2+u_y\right)\,v_1\,dt-v_1\,dx- u_x\,v_1\,dy\right).
\]
This form annules whenever $v$ satisfies the following system of PDEs:
\begin{equation}
v_{t} = \left(\case{1}{2}\,u_x^2+u_y\right)\,v_1,
\qquad
v_{x} =v_1, 
\qquad
v_{y} = u_x\,v_1.
\label{mKhZcovering1}
\end{equation}
Excluding $v_1$ from this system, we have the covering equations 
\begin{equation}
v_{t} = \left(\case{1}{2}\,u_x^2+u_y\right)\,v_{x},
\qquad
v_{y} = u_x\,v_{x}.
\label{mKhZcovering}
\end{equation}
Introducing fibre coordinates $v_0=v$, $v_k = \partial^k\,v/\partial\,x^k$, $k \in \N$, we obtain from (\ref{mKhZcovering1})
the extended total derivatives 
\begin{eqnarray*}
\widetilde{D}_t = \bar{D}_t 
+ \sum \limits_{j =0}^{\infty} \widetilde{D}^j_x(\left(\case{1}{2}\,u_x^2+u_y\right)\,v_1)\,
\frac{\partial}{\partial v_j},
\\
\widetilde{D}_x = \bar{D}_x + \sum \limits_{j =0}^{\infty} v_{j+1}\,\frac{\partial}{\partial v_j},
\\
\widetilde{D}_y = \bar{D}_y 
+ \sum \limits_{j =0}^{\infty} \widetilde{D}^j_x(u_x\,v_1)\,\frac{\partial}{\partial v_j},
\end{eqnarray*}
and the WE forms
\begin{eqnarray*}
\widetilde{\vartheta}_0 = d v_0 - \left(\case{1}{2}\,u_x^2+u_y\right)\,v_1\,dt 
- v_1\,dx- u_x\,v_1\,dy,
\\
\widetilde{\vartheta}_{k} = \widetilde{D}^k_x(\widetilde{\vartheta}_0),
\qquad 
k \in \N.
\end{eqnarray*}
Excluding $u$ from (\ref{mKhZcovering}), we obtain 
\begin{equation}
v_{yy}= v_{tx} +\frac{v_y^2-v_t\,v_x}{v_x^2}\,v_{xx}.
\label{mKhZ_transformed}
\end{equation}
Therefore, (\ref{mKhZcovering}) is a B\"acklund transformation between equations  (\ref{mKhZ}) and (\ref{mKhZ_transformed}).

\section{Conclusion}

We have shown that a covering for a nonlinear PDE in three independent variables can be revealed by means of Cartan's equivalence method. For the modified Khokhlov-Za\-bo\-lot\-ska\-ya equation (\ref{mKhZ}) the covering equations appear from a linear combination of the MC forms of its symmetry pseudo-group. While there is the algorithm for com\-pu\-ting the MC forms, further study is required to enlight the relation between Car\-tan's structure the\-o\-ry of symmetry pseudo-groups and nonlocal aspects of geometry of DEs.

\end{document}